# Puzzle maker in $SmB_6$:

## Accompany-type valence fluctuation state

Qi Wu[1] and Liling Sun[1,2]

[1]*Institute of Physics and Beijing National Laboratory for Condensed Matter Physics, Chinese Academy of Sciences, Beijing 100190, China*

[2]*Collaborative Innovation Center of Quantum Matter, Beijing 100190, China*

**Abstract**

In recent years, the study on the Kondo insulator $SmB_6$, a strongly correlated electron material with decades-long puzzles，has become one of the most attractive topics again because the discovery of the coexistence of its unusual metallic surface state with an insulating bulk [1-6]. Many efforts have been made in understanding the corresponding physics behind in $SmB_6$ [7-10], but some puzzles on it, being hotly debated and argued, has not been solved. In this article, based on the latest progress in our high pressure studies and the accumulating results reported by other groups on $SmB_6$，we propose a notion named as "accompany-type valence fluctuation state", which possibly coexists with the Kondo ground state of $SmB_6$. The purpose of this article is to search a common starting point from which most of the accumulated low-temperature phenomena observed by different experimental investigations on $SmB_6$ could be understood in a unified way. Although this notion is only our personal understanding from a phenomenological point of view and may be immature, anyway, we expect that this notion could attract rigorous theoretical interpretations and further experimental investigations, or stimulate better thinking on the physics in $SmB_6$.

## 1. Introduction

Historically, in the last 70[th], one of the important issues in the field of the strongly correlated electron systems (SCESs) was to reveal the physics through the studies on the materials with mixed-valence state, among which $SmB_6$ was a representative system. Afterwards, the attentions on the SCESs were transferred to the other materials believed to be more attractive, *i.e.* to heavy fermion materials in 1979, to copper oxide superconductors in 1986 and to iron-based superconductors in 2008. However, some fundamental issues on $SmB_6$ had not been solved thoroughly, *e.g.* the origin of the low-temperature resistance plateau (LTRP), understanding the mixed valence behaviors (it is the most prominent feature of $SmB_6$), and so on. New interests in $SmB_6$ are aroused by the recent discoveries of the coexistence of an exotic metallic surface state with a bulk insulating state [10,11] and the understanding for its topological nature of the metallic surface, as well as the unusual quantum oscillation [12,13], all of which push the studies on $SmB_6$ back to the research frontier of the SCESs. However, the convergence of the old puzzles and the new findings makes $SmB_6$ more mysterious. Therefore, a unified understanding on the abundant phenomena observed or discovered in $SmB_6$ by different experimental methods or theories is urgently needed so as to reconcile the interpretations on these results. If a unified understanding is realized, it will be a significant advance in the field of SCESs, because it will not only solve the puzzles in $SmB_6$ but also may be helpful to promote the studies on other strongly correlated electron materials.

The prominent anomaly phenomena in $SmB_6$ are the presence of resistance plateau and its unusual quantum oscillation under magnetic field in almost the same low-temperature range. However, the real ground state responsible for these low-temperature phenomena has yet to be known. Intuitively, the co-emergence of these phenomena is reminiscent of the superconducting state in which the zero resistance and corresponding diamagnetic behaviors occur simultaneously. The responses of the superconducting ground state to the applied current or magnetic field lead us to raise the question on $SmB_6$ whether all puzzling low-temperature behaviors observed by applying current or magnetic field also share the same physical origin, *i.e.* a peculiar ground state unknown. In this article, we propose that the ambient-pressure $SmB_6$ is in a bulk Kondo-singlet ground state with a unique valence fluctuation, and tentatively classify this valence fluctuation as the "accompany-type valence fluctuation state" (AVFS for short), which is expected to be helpful in deciphering the puzzles in SmB$_6$.

## 2. Lower and upper valence limits for formation of resistance plateau

Our motivation to clarify the ground state of SmB$_6$ was kindled by our high pressure studies on it [14,15]. It is believed that pressure is a 'clean' way of tuning the crystal and electronic structures without introducing additional chemical complexity into the studied system [16,17]. A peculiar effect of pressure on the rare earth hexaborides REB$_6$ (RE=Sm and Yb) is that it can increase the mean valence of the rare earth ions and enhance the population of magnetic ions accordingly. Our recently high pressure studies on the relationship between the mean valence and low-temperature properties in SmB$_6$ reveal that its ground state is closely connected with the instability of the mixed valence state [14,15]. It is found that pressure can simultaneously tune both the onset temperature of the LTRP and the mean valence in SmB$_6$. The resistance plateau is found to disappear at ~ 4 GPa, meanwhile its mean valence shows an increases from 2.54 at ambient pressure to 2.62 at ~4 GPa. Intriguingly, the LTRP is also discovered in the pressurized YbB$_6$, a sister compound of SmB$_6$, at the pressure above 15 GPa where its mean valence increases to 2.03 from 2.00 at ambient pressure [16].

From the above high-pressure experimental results, we know that there exist an upper valence limit and a lower valence limit for the formation of the LTRP in the rare earth hexaboride system REB$_6$ (RE=Sm, Yb). The upper limit (2.62$^+$) presents in the pressurized SmB$_6$, while the lower limit (2.03$^+$) shows in the pressurized YbB$_6$, as shown in Fig.1. Consequently, an important question whether an appropriate mixed valence state is the necessary condition for developing the LTRP in REB$_6$ is raised. It is known that SmB$_6$ and YbB$_6$ possess the same CsCl-type lattice structure constructed by the B$_6$ framework and the rare earth ions located in the interstitial of the framework. The interstitial space is highly symmetric and large enough to host all kinds of rare earth elements from Ce to Lu to form all kinds of rare earth hexaborides. More interestingly, SmB$_6$ and YbB$_6$ are the only two hexaborides who contain the rare earth elements with the configuration balance of $4f^n$(NM)–$4f^{n-1}$(M)$5d^1$ (here NM stands for non-magnetic and M stands for magnetic, and the total angular momentum $J$ for $4f^n$ and $4f^{n-1}$ are non-integer and integer respectively) or $Sm^{2+}/Yb^{2+}$ →$Sm^{3+}/Yb^{3+}$+$5d$ correspondingly [18], as shown in Table 1. This indicates that the populations of their magnetic $4f$ electrons and $5d$ electrons can increase or decrease together with the change of pressure. This pressure effect on the populations of magnetic ions and conduction electrons definitely impact on the peculiar metallic surface state and the bulk insulating state differently in the systems and consequently generate abundant of physics phenomena.

## 3. Role of temperature-induced valence change in developing resistance plateau

Recent high pressure studies on $YbB_6$ by Zhou *et al* demonstrate a pressure-induce valence change of Yb ions from $2^+$ (ambient pressure) to $2.09^+$ (20 GPa), just where a clear resistance plateau is observed [16]. Can such a minor valence change in $YbB_6$ really play a vital role in the development of the resistance plateau? The positive answer can be drawn from the delicate valence measurements performed by Masaichiro *et al* at ambient pressure and different temperatures [19], as shown in Fig.2. It can be seen that the mean valence ($v$) of Sm ions decreases from 2.59 at room temperature to 2.52 at 2 K, where the resistance plateau presents. The crucial importance of this temperature-induced valence change in developing the LTRP behaviors of ambient-pressure $SmB_6$ can be reasonably understood by comparing it with the high pressure case. The value of this temperature-induced mean valence variation ($\Delta v = v_{300K} - v_{2K} = 0.07$) in $SmB_6$ is comparable with that of the pressure-induced mean valence change ($\Delta v = v_{15GPa} - v_{28GPa} = 0.08$) in $YbB_6$ for the development of the resistance plateau.

## 4. Accompany-type valence fluctuation state

Evoked by the correlation between the formation of resistance plateau and the upper/lower valence limits, as well as the temperature-induced valence instability at ambient pressure and a plenty of related results reported (especially the existence of magnetic fluctuation [20], charge fluctuation [21] and valence fluctuation [22-24] from a diverse experimental methods), we propose that the ambient-pressure $SmB_6$ is in a peculiar valence fluctuation state coexisting with Kondo insulating state, which is different from that of other rare earth hexaborides. Here, we tentatively classify the valence fluctuation state as the "accompany-type valence fluctuation state" (AVFS for short). Some main points on the AVFS can be summarized as follows:

**1) Notion** of the AVFS is distinguished by the peculiar fashion of the valence fluctuation state, in which the magnetic ions and $d$ electrons increase or decrease together, as shown in Fig.3a. It should be noted that this type of valence fluctuation can exist only in the Sm or Yb-containing hexaborides. While, for the other rare earth elements with both divalent and trivalent states, such as Eu, the population's change of the magnetic ions and $d$ electrons can be described as one fall and the other raise in the valence fluctuation state, as if the $f$ electron determining the magnetic state of Sm ion and the $5d$ electron transfers each other, as shown in Fig.3b. Therefore, the term of "transfer-type valence fluctuation state" (TVFS for short) can be adopted for this case for comparison and the convenience of further discussion.

**2) Necessary condition** for developing LTRP in the $REB_6$ system is that the compound is required to have an appropriate value of mean valence. Therefore, a suitable value of the mean valence may be also a possible necessary condition for the formation of AVFS, which deserves further studies.

**3) Support evidence** for existence of the AVFS in $SmB_6$ include i) the existence of magnetic, charge and valence fluctuations [20-23]; ii) observation of a self-sustained voltage oscillation [25], which should be associated with the AVFS; iii) different ratios of $Sm^{2+}$ and $Sm^{3+}$ (0.6-0.7:0.4-0.3) [22], as well as its dependence on temperature [19], all of which should be intimately connected with the instability associated with the AVFS.

## 5. Primary analysis on the mechanism

Generally, the possible mechanism of the proposed AVFS in such a Kondo system is originated from the interplay between the mixed-valence Sm ions with a unique configuration of the $f$ electrons ($4f^n$(NM)–$4f^{n-1}$(M)$5d^1$) and the special lattice of the $B_6$ framework which provides an appropriate environment to stabilize the AVFS. Nevertheless, it is known that the finding of a new state of matter and the uncovering of its mechanism usually are of different knowing stages for the host material, *i.e* primary stage and final stage. Such as we can confirm a superconducting ground state in a high Tc superconductor by testing both of its zero resistance and diamagnetism, however fully understanding its superconducting mechanism has not been achieved, although after decades' efforts, and is still one of the 21 century's challenges to the field of condensed matter physics. In the case of $SmB_6$, if the AVFS state can be confirmed by further studies, it may or may not need such a long time for understanding its mechanism. Supposing that the AVFS can be confirmed by the experimental criteria of observing the LTRP and the anomaly quantum oscillation, *i.e.* the responses to electric field and magnetic field respectively, consequently what the mechanism for the AVFS is should be a new challenge to the field of strongly correlated electron systems.

Here, we try to give a primary explanation for the existence of AVFS based on our complimentary analysis on the related experimental results on $SmB_6$. We propose that the AVFS is a kind of quantum oscillation between two different states (corresponding to $4f^n$ and $4f^{n-1}5d^1$ states in $SmB_6$), and suggest that this quantum oscillation may exist in such a system as $SmB_6$, in which the Sm ion protected by $5s$ and $5p$ shell is trapped in the interstitial of $B_6$ framework. The interstitial is highly symmetric and can provide the required environment for the AVFS. The in-depth discussion about the possible mechanism of the AVFS is beyond the scope of this article.

As a matter of fact, we noted that the concept of valence fluctuation in $SmB_6$ was proposed primitively by Kasuya *et al* in 1979 [22], in which it was even pointed out that the valence is fluctuating in the time scale with a frequency between $10^{-9}$ and $10^{-15}$s, probably around $10^{-13}$s. However, it was lack of precise descriptions on the state and especially they inappropriately believed that the ratio of the $Sm^{2+}$ and $Sm^{3+}$ is unchanged with temperature. Also, we noted that most of the recent studies on $SmB_6$ did mention the mixed valence feature, but they are lack of an appropriate consideration on its AVFS.

## 6. Application of AVFS notion for studies on $SmB_6$

Currently, the most attractive issue on studies of $SmB_6$ is the interpretation on its exotic metallic surface state. Here, let us have a try to make a rough analysis on the origination of the metallic surface from the perspective of the AVFS notion, and also give an intuitive explanation about different effects of the AVFS and the TVFS on the surface of the $SmB_6$ sample. As mentioned above, the existence of the AVFS requires that the interstitial of the $B_6$ framework is highly symmetric to provide an appropriate environment to generate the AVFS. While on its surface the lattice symmetry is broken and the conduction is better than the bulk due to the existence of Boron's *p* dangling bonds and the influence of Rashba effect [26,27], thus populations of conduction electrons and magnetic ions are higher than the bulk. As a result, the AVFS cannot be survived on the surface. Just as Coleman pointed out that, on the ground state of $SmB_6$, the surface Kondo singlet breaks down [3]. This breakdown is supported indirectly by our high pressure studies on $SmB_6$. We found that at pressure above ~ 4 GPa the mean valence is increased to a similar level as what is observed in ambient-pressure $SmB_6$ at very low temperature [14]. Simultaneously the bulk Kondo state is collapsed and the LTRP (signifying the metallic surface state) disappears [14].

In addition, we argue that the TVFS cannot give rise to a metallic surface state as the AVFS does in $REB_6$, because in the TVFS the population of magnetic rare earth ions on the surface will decrease with the increasing of the 5*d* electrons. This change is in favor of stabilizing the Kondo singlet.

Besides the issues discussed above, there are many other interesting questions on $SmB_6$ deserving to be further studied from experimental and theoretical sides, such as the interpreting the correlation between the AVFS and the magnetic field-induced unusual quantum oscillation in an insulator; the effects of the AVFS on some other oscillation phenomena [25]; the interplays between the AVFS and the surface dangling bonds of Boron's *p* electrons [26] which is a common feature of rare earth hexaborides; the relationship between the AVFS and the Kondo effect or Rashba effect [27]; the connection of the AVFS with negative volume shrinking upon temperature decrease [25,28] as well as with the coexistence and interplay of the different valence states between surface and bulk [29].

However, it is noteworthy that, even if the AVFS idea or other better scenario can be adopted as a common starting point for the studies on $SmB_6$, the different or even conflicting conclusions might unavoidably be achieved because of the multiple entangled factors interacting sensitively with the exotic metallic surface state in such a complicated Kondo system，which may yield different results observed and lead to diverse conclusions. These factors can be (1) samples prepared by different growing methods and techniques, which can give different growing surfaces and different qualities for the sample (atom vacancies, impurities and their magnetic ion population etc.). Here we should mention that a great deal of high quality samples of $SmB_6$ synthesized by Fisk's group have been broadly used in many important experiments for decades; (2) the difference of the sensitivity of detecting techniques for the surface and bulk state, including the differences in testing methods and the parameters adopted for the testing by the similar equipment. For example, the different frequencies (energies) of the excitation sources adopted in an angle-resolved photoemission electronic spectroscopy (ARPES) or in the quantum oscillation measurements may collect different proportions of information from the surface and the bulk state; (3) the different detecting methods may also give rise to different volume ratio between the surface state and the bulk state of the sample [30].

In summary, we propose that $SmB_6$ is a special kind of Kondo insulator with a bulk

"accompany-type valence fluctuations" which is prohibited on the surface. As a result, the Kondo singlet on the surface breaks down and the corresponding anomaly metallic surface state disappears. If the notion of the AVFS is adoptable as a starting point for the studies on the possible topological Kondo insulator $SmB_6$, it can be optimistically expected that a more unified or precise understanding could be achieved, including the origination of the puzzling LTRP, the magnetic field-induced unusual quantum oscillation and the topological nature of the metallic surface etc..

## Acknowledgments

We thank M. Aronson, P. Coleman, Z. Fisk, L. H. Greene, Q. M. Si, J. D. Thompson, Y. F Yang, R. Yu, G.M. Zhang, Z. X. Zhao for helpful discussions. Authors are grateful to Y. Z. Zhou for his assistance in reference edit. The work was supported by the NSF of China (Grants No. 91321207, No. 11427805, No. U1532267, No. 11404384), the Strategic Priority Research Program (B) of the Chinese Academy of Sciences (Grant No. XDB07020300), and the National Key Research and Development Program of China (Grant No.2016YFA0300300).

To whom correspondence should be addressed.
E-mail: llsun@iphy.ac.cn and wq@iphy.ac.cn

## References

[1] Denlinger J D, Allen J W, Kang J S, Sun K, Min B I, Kim D J, and Fisk Z, in *Proceedings of the International Conference on Strongly Correlated Electron Systems (SCES2013)* (Journal of the Physical Society of Japan, 2014).

[2] Kim D J, Xia J, and Fisk Z 2014 Topological surface state in the Kondo insulator samarium hexaboride *Nat. Mater.* **13** 466

[3] Erten O, Ghaemi P, and Coleman P 2016 Kondo Breakdown and Quantum Oscillations in $SmB_6$ *Phys. Rev. Lett.* **116** 046403

[4] Nakajima Y, Syers P, Wang X, Wang R, and Paglione J 2016 One-dimensional edge

state transport in a topological Kondo insulator *Nat. Phys.* **12** 213

[5] Xu Y, Cui S, Dong J K, Zhao D, Wu T, Chen H X, Sun K, Yao H and Li S Y 2016 Bulk Fermi surface of charge-neutral excitations in SmB$_6$ or not: a heat-transport study arXiv 1603 09681

[6] Park W K, Sun L, Noddings A, Kim D-J, Fisk Z, and Greene L H 2016 Topological surface states interacting with bulk excitations in the Kondo insulator SmB$_6$ revealed via planar tunneling spectroscopy, *PNAS*, **113** 6599

[7] Alexandrov V, Dzero M, and Coleman P 2013 Cubic Topological Kondo Insulators *Phys. Rev. Lett.* **111** 226403

[8] Lu F, Zhao J, Weng H, Fang Z, and Dai X 2013 Correlated Topological Insulators with Mixed Valence *Phys. Rev. Lett.* **110** 096401

[9] Zhang X, Butch N P, Syers P, Ziemak S, Greene R L, and Paglione J 2013 Hybridization, Inter-Ion Correlation, and Surface States in the Kondo Insulator SmB$_6$ *Phys. Rev. X* **3** 011011

[10] Xu N, Matt C E, Pomjakushina E, Shi X, Dhaka R S, Plumb N C, Radović M, Biswas P K, Evtushinsky D, Zabolotnyy V, Dil J H, Conder K, Mesot J, Ding H, and Shi M 2014 Exotic Kondo crossover in a wide temperature region in the topological Kondo insulator **SmB$_6$** revealed by high-resolution ARPES *Phys. Rev. B* **90** 085148

[11] Neupane M, Alidoust N, Xu S Y, Kondo T, Ishida Y, Kim D J, Liu C, Belopolski I, Jo Y J, Chang T R, Jeng H T, Durakiewicz T, Balicas L, Lin H, Bansil A, Shin S, Fisk Z, and Hasan M Z 2013 Surface electronic structure of the topological Kondo-insulator candidate correlated electron system SmB$_6$ *Nat. Commun.* **4** 7

[12] Li G, Xiang Z, Yu F, Asaba T, Lawson B, Cai P, Tinsman C, Berkley A, Wolgast S, Eo Y S, Kim D-J, Kurdak C, Allen J W, Sun K, Chen X H, Wang Y Y, Fisk Z, and Li L 2014 Two-dimensional Fermi surfaces in Kondo insulator SmB$_6$ *Science* **346** 1208

[13] Tan B S, Hsu Y-T, Zeng B, Hatnean M C, Harrison N, Zhu Z, Hartstein M, Kiourlappou M, Srivastava A, Johannes M D, Murphy T P, Park J-H, Balicas L, Lonzarich G G, Balakrishnan G, and Sebastian S E 2015 Unconventional Fermi surface in an insulating state *Science* **349** 287

[14] Zhou Y, Wu Q, Rosa P F S, Yu R, Guo J, Yi W, Zhang S, Wang Z, Wang H, Cai S, Yang

K, Li A, Jiang Z, Zhang S, Wei X, Huang Y, Yang Y-f, Fisk Z, Si Q, Sun L, and Zhao Z 2016 Quantum phase transition and destruction of Kondo effect in pressurized SmB$_6$ arXiv:1603 05607

[15] Sun L and Wu Q 2016 Pressure-induced exotic states in rare earth hexaborides *Rep. Prog. Phys.* **79** 084503

[16] Zhou Y, Kim D-J, Rosa P F S, Wu Q, Guo J, Zhang S, Wang Z, Kang D, Yi W, Li Y, Li X, Liu J, Duan P, Zi M, Wei X, Jiang Z, Huang Y, Yang Y-f, Fisk Z, Sun L, and Zhao Z 2015 Pressure-induced quantum phase transitions in a YbB$_6$ single crystal *Phys. Rev. B* **92** 241118

17. Park T, Ronning F, Yuan H Q, Salamon M B, Movshovich R, Sarrao J L & Thompson J D.2006 Hidden magnetism and quantum criticality in the heavy fermion superconductor CeRhIn$_5$. *Nature* **440**, 65

[18] Blundell S 2001 Magnetism in Comdensed Matter *Oxford University press* page 34

[19] Masaichiro M, Satoshi T, and Fumitoshi I 2009 Temperature dependence of Sm valence in SmB$_6$ studied by X-ray absorption spectroscopy *Journal of Physics: Conference Series* **176** 012034

[20] Biswas P K, Salman Z, Neupert T, Morenzoni E, Pomjakushina E, von Rohr F, Conder K, Balakrishnan G, Hatnean M C, Lees M R, Paul D M, Schilling A, Baines C, Luetkens H, Khasanov R, and Amato A 2014 Low-temperature magnetic fluctuations in the Kondo insulator SmB$_6$ *Phys. Rev. B* **89** 161107

[21] Min C-H, Lutz P, Fiedler S, Kang B Y, Cho B K, Kim H D, Bentmann H, and Reinert F 2014 Importance of Charge Fluctuations for the Topological Phase in **SmB$_6$** *Phys. Rev. Lett.* **112** 226402

[22] Kasuya T, Takegahara K, Fujita T, Tanaka T, and Bannai E 1979 Valence Fluctuating State in SmB$_6$ *J. Phys. Colloques* **40** C5

[23] Takigawa M, Yasuoka H, Kitaoka Y, Tanaka T, Nozaki H, and Ishizawa Y 1981 NMR Study of a Valence Fluctuating Compound SmB$_6$ *J. Phys. Soc. Jpn.* **50** 2525

[24] Ruan W, Ye C, Guo M, Chen F, Chen X, Zhang G-M, and Wang Y 2014 Emergence of a Coherent In-Gap State in the SmB$_6$ Kondo Insulator Revealed by Scanning

Tunneling Spectroscopy *Phys. Rev. Lett.* **112** 136401

[25] Kim D J, Grant T, and Fisk Z 2012 Limit Cycle and Anomalous Capacitance in the Kondo Insulator **SmB$_6$** *Phys. Rev. Lett.* **109** 096601

[26] Zhu Z H, Nicolaou A, Levy G, Butch N P, Syers P, Wang X F, Paglione J, Sawatzky G A, Elfimov I S, and Damascelli A 2013 Polarity-Driven Surface Metallicity in SmB$_6$ *Phys. Rev. Lett.* **111** 216402

[27] Sengupta P, Matsubara M, and Bellotti E 2015 The intrinsic X point longitudinal conductivity and Berry phase of the topological Kondo insulator SmB$_6$ arXiv:1509.

[28] Curnoe S and Kikoin K A 2000 Electron self-trapping in intermediate-valent SmB$_6$ *Phys. Rev. B* **61** 15714

[29] Phelan W A, Koohpayeh S M, Cottingham P, Freeland J W, Leiner J C, Broholm C L, and McQueen T M 2014 Correlation between Bulk Thermodynamic Measurements and the Low-Temperature-Resistance Plateau in SmB$_6$ *Phys. Rev. X* **4** 031012

[30] Syers P, Kim D, Fuhrer M S, and Paglione J 2015 Tuning Bulk and Surface Conduction in the Proposed Topological Kondo Insulator SmB$_6$ *Phys. Rev. Lett.* **114** 096601

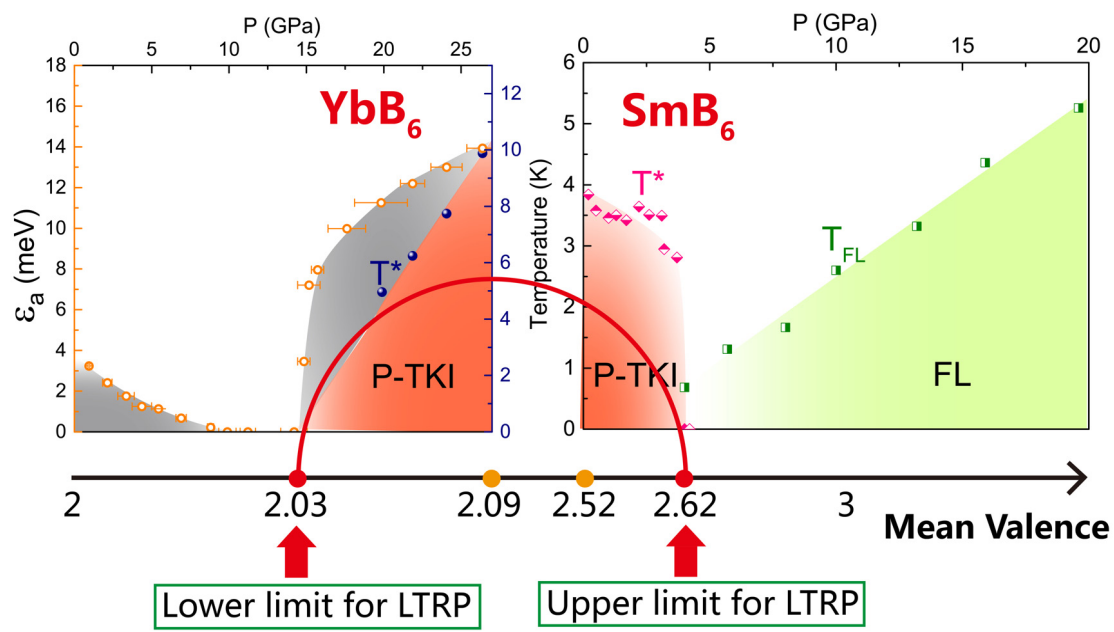

Fig.1 Combined phase diagram of pressure dependence of turn-on temperatures (T*) of the resistance plateaus in YbB$_6$ (left panel) and SmB$_6$ (right panel), and the relation with their mean valence. P-TKI and FL stand for the putative topological Kondo insulating state and Fermi liquid state, respectively. T$_{FL}$ in the right panel represents Fermi liquid temperature. $\varepsilon_\alpha$ in the left panel is activation energy obtained by fitting to the resistance-temperature data. The red arrows indicate the lower and upper limits of the mean valence for the development of the low-temperature resistance plateau in REB$_6$ (RE=Sm,Yb).

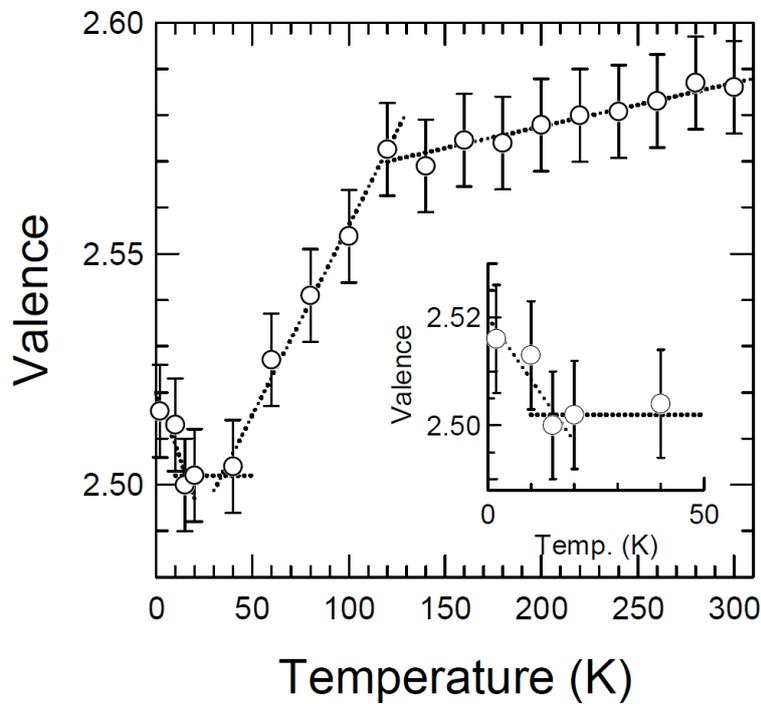

Fig. 2 Temperature dependence of the mean valence of Sm ions in SmB$_6$ (the figure is taken from Ref. [19] with permission from the publisher). The low-temperature upturn indicates that the population of the Sm ions with tri-valence state is increased as the temperature decreases and may feature the formation of the AVFS.

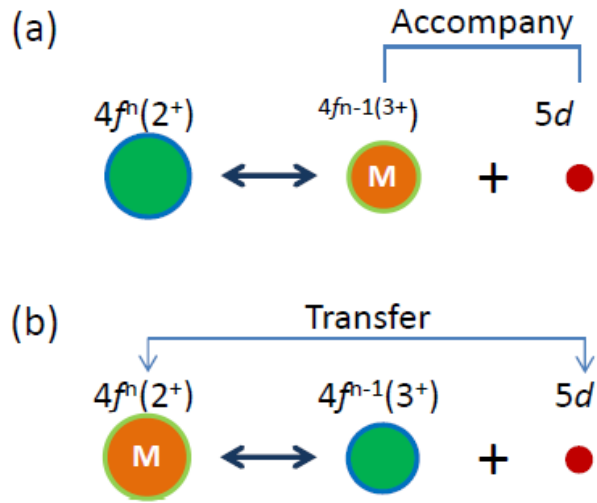

**Fig. 3** A schematic description on the difference of the relations among the *f* electron configuration, magnetic ion and *d* electron between (a) the accompany-type valence fluctuation and (b) the transfer-type valence fluctuation.

**Table 1** The magnetic ground states of the 4f ions for the rare earth elements with di-valence state. For each ion, by using Hund's rules, the shell configuration and the predicted values of S (spin momentum), L (orbital momentum) and J (total angular momentum), as well as the term symbol of $^{2s+1}L_J$, are given. The data derived from [18]. The electron configurations and the corresponding terms in red color indicate the ion in a magnetic state. It can be seen that, when the valence is increased from $2^+$ to $3^+$, the electron configuration of Sm or Yb changes from non-magnetic state to magnetic state, while the electron configuration of Eu or Tm changes from magnetic state to non-magnetic state.

| Ion | Shell | S | L | J | term |
|---|---|---|---|---|---|
| Sm$^{2+}$ (Sm$^{3+}$) | $4f^6$($4f^5$) | 3(5/2) | 3(5) | 0(5/2) | $^7F_0$($^6I_{5/2}$) |
| Eu$^{2+}$ (Eu$^{3+}$) | $4f^7$($4f^6$) | 7/2(3) | 0(3) | 7/2(0) | $^8S_{7/2}$($^7F_0$) |
| Tm$^{2+}$ (Tm$^{3+}$) | $4f^{13}$($4f^{12}$) | 1/2(1) | 3(5) | 7/2(6) | $^2F_{7/2}$($^3H_6$) |
| Yb$^{2+}$ (Yb$^{3+}$) | $4f^{14}$($4f^{13}$) | 0(1/2) | 0(3) | 0(7/2) | $^1S_0$($^2F_{7/2}$) |